# SELFISH CARRIER MONITORING IN WIFI USING DISTRIBUTED SNIFFERS

U.Sinthuja[1] M.Phil. (Research Scholar),
R.Sridevi[2] (Asst. Professor)
Department of Computer Science, PSG College of Arts & Science, Coimbatore, India

___________________________________________________________________________

***Abstract***: *This work proposes a tool to estimate the interference between nodes and links in a live wireless network by passive monitoring of wireless traffic. This approach requires deploying multiple sniffers across the network to capture wireless traffic traces. These traces are then analyzed using a machine learning approach to infer the carrier-sense relationship between network nodes. It also demonstrates an important application of this tool-detection of selfish carrier-sense behavior. This is based on identifying any asymmetry in carrier-sense behavior between node pairs and finding multiple witnesses to raise confidence. Simulation results demonstrate that the proposed approach of     estimating interference relations is significantly more accurate than simpler heuristics and quite competitive with active measurements. Minimizing router overhead taken as a main goal.*

___________________________________________________________________________

## I. Introduction

The term wireless refers to computers that can communicate with each other without using any wire. Unlike LAN which connects computers with kind of cabling like UTP (Unshielded twisted pair); in wireless network, no data cabling is required. The users in this type of network can share data files and other resources without any requirement to connecting to each other physically. The noticeable advantages of a wireless network are easily seen when considering the needs of users of mobile devices, i.e. handheld PC's, mobile phones and laptops. The term Wi-Fi (Wireless Fidelity) defined as a wireless networking technology which works with no physical connection between sender and receiver by using radio frequency (RF) technology. The term *Wi-Fi* is often used as a synonym for IEEE 802.11 technology. The access areas which provide Internet access through wireless local area network (WLAN) are called "**hot spots**".

Interference is a phenomenon in which two waves superimpose to form a resultant wave of greater or lower amplitude. Interference usually refers to the interaction of waves that are correlated or coherent with each other, either because they come from the same source or because they have the same or nearly the same frequency.  In communications and electronics, especially in telecommunications, interference is anything which alters, modifies, or disrupts a signal as it travels along a channel between a source and a receiver.

Selfish node is a kind of attacker node. This can collect the details of neighbor nodes but it cannot send its detail to neighbor nodes. So here risk finding out the attacker to keep safe our data to be send. One advantage of asymmetric behavior is, nodes in network can sends acknowledgement to sender node if it not receives the data packet. But in symmetric behavior this cannot be done. That's the main reason to using the asymmetric behavior in this project. Because the selfish node cannot sends detail it may after receives the data packet.

## II. Routing Protocols in Network

A **routing protocol** specifies how routers communicate with each other, disseminating information that enables them to select routes between any two nodes on a computer network.

### A. Proactive

The nodes maintain a table of routes to every destination in the network, for this reason they periodically exchange messages. At all times the routes to all destinations are ready to use and as a consequence initial delays before sending data are small. Keeping routes to all destinations up-to-date, even if they are not used, is a disadvantage with regard to the usage of bandwidth and of network resources.

*i) DSDV (Destination-Sequence Distance Vector) Routing*

DSDV has one routing table, each entry in the table contains: destination address, number of hops toward destination, next hop address. Routing table contains all the destinations that one node can communicate.

**Figure -1 Routing Protocols Overview.**

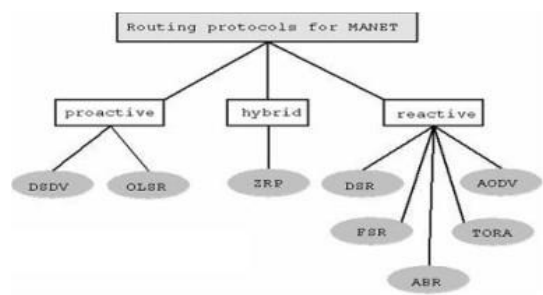

### B. Reactive

These protocols were designed to overcome the wasted effort in maintaining unused routes. Routing information is acquired only when there is a need for it. The needed routes are calculated on demand. This saves the overhead of maintaining unused routes at each node, but on the other hand the latency for sending data packets will considerably increase.

*i) AODV (Ad hoc on demand distance vector) Routing*

AODV routing is the combination of DSDV and DSR. In AODV, each node maintains one routing table. Each routing table entry contains:
- Active neighbor list: a list of neighbor nodes that are actively using this route entry. Once the link in the entry is broken, neighbor nodes in this list will be informed.
- Destination address
- Next-hop address toward that destination
- Number of hops to destination
- Sequence number: for choosing route and prevent loop
- Lifetime: time when that entry expires

*ii) DSR (Dynamic Source Routing)*

DSR is a reactive routing protocol which is able to manage a MANET without using periodic table-update messages like table-driven routing protocols do. DSR was specifically designed for use in multi-hop wireless ad hoc networks.

*iii) TORA (Temporary Ordered Routing Algorithm)*

TORA is based on link reversal algorithm. Each node in TORA maintains a table with the distance and status of all the available links. TORA has three mechanisms for routing:
- Route Creation
- Route Maintenance
- Route Erasure

### III. Previous Approach

A. Problem of Previous

In 802.11, interference can occur either at the "sender side" or at the "receiver side" (or both). Sender side interference pertains to deferral due to carrier sensing. In this case, one node freezes its back off counter and waits when it senses the second node's transmission. In case of receiver side interference, overlapped packet transmission causes collisions at the receiver. This requires packet retransmission. In both cases, the sender additionally has to go through a back off period, when the medium must be sensed idle. The net effect of the interference is reduction of throughput capacity of the network.

General goal is to understand the deferral behavior that accounts for the sender side interference. To detect selfish carrier-sense behavior, we need to identify the asymmetry in the deferral behavior. The deferral behavior between two nodes, X and Y is said to be asymmetric if Y defers for X's transmission and X does not defer for Y's, or vice versa. Such asymmetry is possible in wireless networks due to interface heterogeneity. But it is simply unlikely that a node X demonstrates similar asymmetry with many such Y's in the same direction. Our strategy is to flag such nodes as potentially selfish, with degree of selfishness indicated by extent of asymmetries exhibited and the number of such Y's (called "witnesses").

For modeling convenience, we consider interference between node or link pairs only. Note that it will allow us to capture the "physical interference" [26] where a given link is interfered collectively by a set of other links, not by a single link alone. This is due to the additive nature of the received power. However, pair wise consideration can still be quite powerful in practice. This simplification is not fundamental to our basic technique. The technique can be extended, albeit with higher computational cost, to physical interference.

B. Discussions

To estimate the interference relations between a given pair of nodes, our technique needs to have instances when simultaneous transmissions are attempted by the two nodes. The conjecture here is that if one observes the live network traffic for a long enough periods, enough of such instances will be available for each node pair.

Main target is to
1) Identify such instances,
2) Infer the deferral behaviors during such instances.

There are several challenges here. First, creating a complete and accurate trace is itself a difficult problem. There are many approaches proposed in literature to create a complete trace. But in technique, incomplete trace may suffice as long as it is statistically similar to the complete trace.

### IV. Proposed Approach

Our basic approach is as follows: we model the 802.11 MAC-layer operations of two sender nodes in the network (say, X, Y) via a Markov chain. The parameters of this chain (essentially the state transition probabilities) are estimated from the observed trace using an approach based on the Hidden Markov Model (HMM) [27].

### A. Hidden Markov Model for Sender-Side Interactions

A hidden Markov model [27] represents a system as a Markov chain with unknown parameters. Here the states of the Markov chain are not directly visible, but some observation symbols influenced by the states are visible. The unknown parameters (such as the state transition probabilities of the Markov chain) can be learned using different standard methods [27], [28], [29] with the help of the observed sequence of observation symbols. Various machine learning applications such as pattern, speech, and handwriting recognition have used HMM technique. We will be using the HMM approach for modeling interactions between a pair of senders in an 802.11 network and inferring sender-side interference relations (deferral behavior) between them.

**Figure-2 Markov model of the combined MAC Layer behavior of two nodes (Sender side only)**

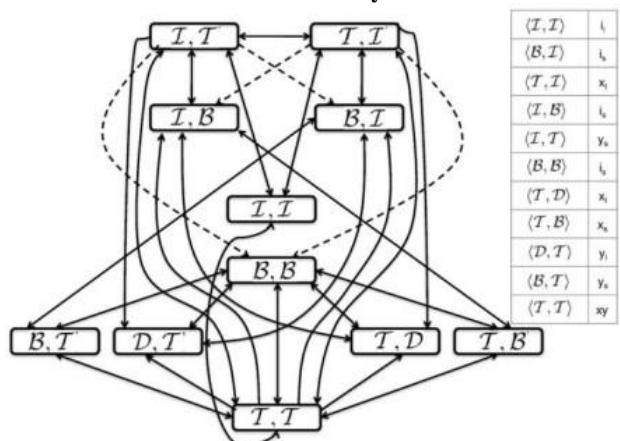

In above figure for the two-node combined Markov chain. (Only the solid lines indicate valid transitions). Here, each state is a tuple consisting of states of individual nodes. Such a Markov chain would be intractable as it would lead to a state space explosion with exponential number of states. Markov chain are not directly visible in the packet trace. Instead a set of observation symbols are visible. There are four possible observation symbols in the trace depending on whether X or Y transmits:

. i: neither X, nor Y transmitting.
. x: X transmitting.
. y: Y transmitting.
. xy: both X and Y transmitting.

### B. Detection of Selfish and Malicious Behavior

The selfish carrier-sense behavior using the pair wise interference relationships discovered by the proposed technique. A set of "sniffers" are deployed to collect traffic traces from a live network. The traffic traces are then merged using existing merging techniques for distributed sniffer traces. Here a machine learning-based approach to analyze the merged traces to infer sender-side interference relationship. Here detect the selfishness and the malicious behavior by using the passive monitoring system.

*i) Detecting Asymmetric Behavior*

To detect selfish carrier-sense behavior, we need to identify asymmetric behavior. This can be detected using the following fashion. The probability that X has a packet to transmit and it defers while Y transmits. The opposite probability (i.e., Y has a packet to transmit and it defers while X transmits) Larger the difference, higher is the asymmetry. Due to the nature of our approach, the asymmetry is tested between a node pair at a time.

*ii) Witness choosing*

In general, each network node X must be evaluated for selfish behavior. By default, every other node Y acts as a witness and the above metric of asymmetry is evaluated for the pair (X,Y) Thus, for each network node X, Here take the average of the metric of asymmetry n(X,Y) over all the witnesses Y that provide a positive value. The negative values are discounted as they will be accounted when Y is evaluated with X as the witness. This average is "selfishness metric."

### C. Simulation

Network simulator (NS) is an object–oriented, discrete event simulator for networking research. Network simulator 2 is used as the simulation tool in here.

Ns2 simulations let us implement various degrees of selfishness, where the selfish node senses carrier with only a certain probability. We use the term degree of selfishness $P_8$ to indicate that the selfish node senses carrier with probability equal to 1-$P_8$. Ns2 simulations also make it easier to investigate larger networks, where there are many nodes, possibly with more than one selfish node with varying traffic and degrees of selfishness.

## IV. Minimizing Router Overhead

In the proposed work the routing path is founded by the sender side and if there is any in this modification work, to reduce the routing overhead and to improve the security in the WI-FI transmissions. In this work already suppressed the interference problems and we activate the passive monitoring system even when the network is in connected state or disconnected state. By reducing the routing overhead problem also reduce the delay and can improve the performance of the network.

The following considerations are taken to be router overhead reduction,
- Packet Drop
- Packet Delivery Ratio
- Delay

### A. Throughput

Network Throughput refers to the volume of data that can flow through a network. Network Throughput is constrained by factors such as the network protocols used, the capabilities of routers and switches, and the type of cabling, such as Ethernet and fiber optic, used to create a network. Network Throughput in wireless networks is constrained further by the capabilities of network adapters on client systems.

$$THROUGHPUT == NO.OF\ BYTES * 8.0 / INTERVAL\ TIME * 1000$$

### B. Packet delivery ratio

The packet delivery ratio can be defined as total number of packets received by total number of packets sent.

$$PDR = TOTAL\ PACKET\ RECEIVED\ /\ TOTAL\ PACKET\ SENT$$

### C. Delay

Network delay is an important design and performance characteristic of a computer network or telecommunications network. The delay of a network specifies how long it takes for a bit of data to travel across the network from one node or endpoint to another. It is typically measured in multiples or fractions of seconds. Delay may differ slightly, depending on the location of the specific pair of communicating nodes. Although users only care about the total delay of a network, Transmission delay is a function of the packet's length and has nothing to do with the distance between the two nodes. This delay is proportional to the packet's length in bits,

$$DELAY = N/R$$

Where
  N is the number of bits, and
  R is the rate of transmission (say in bits per second)

## V. Simulation Discussions

Simulations can create arbitrary topologies and interference conditions easily. However, the physical layer (including interface behavior for carrier sense and packet capture) implementation is often idealized or unrealistic in simulations. To address this issue, extended version of the NS2 simulator that includes realistic measurement- based models. These models were validated against experimental results showing excellent accuracy. In all following automatically generated graphs while simulation remains X axes having the values of time and the Y axes denotes the values of number of packets x10.

In figure-3 red color shows the results of existing system and the green color line shows the results after minimizing the router overhead problem.Figure-4 shows the accuracy of delay results shows the values after the minimization of router overhead. Red in color denotes the existing and the green in color determines the proposed values. Minimizing the work of router overhead reduces the delay.

**Figure-3 Packet Delivery Ratio**

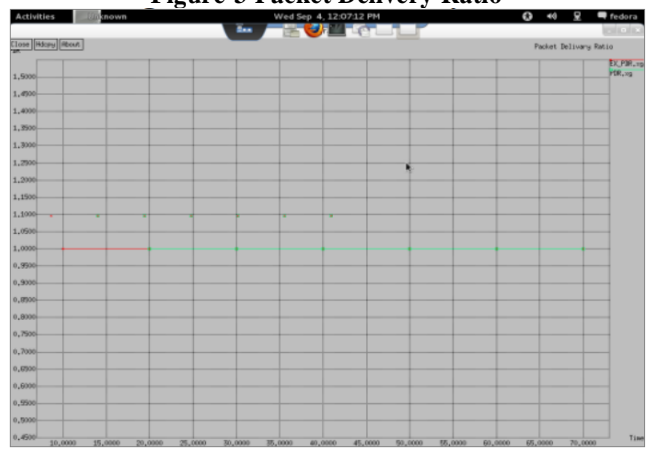

### Figure-4 Packet Delay

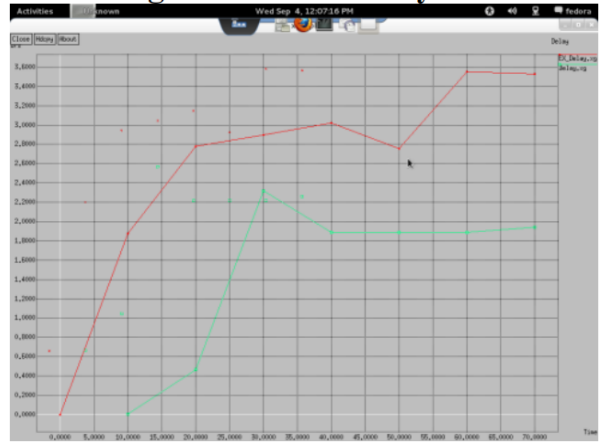

### VI. Conclusion & Future Focus

Here investigated a novel machine learning-based Approach to estimate interference and to detect selfish carrier-sense behavior in an 802.11 network. The technique uses a merged packet trace collected via distributed sniffing. It then recreates the MAC layer interactions on the sender side between network nodes via a machine learning approach using the Hidden Markov Model. This coupled with an estimation of collision probability on the receiver side is helpful in inferring the probability of interference in the network links. Significant asymmetry in the sender -side interaction in favor of a particular node witnessed by multiple other nodes indicates selfishness. The power of this technique is that it is purely passive and does not require any access to the network nodes. The comparison simulation results indicates the minimizing the router overhead reduces the delay of packets.

Our future work will include more evaluations to demonstrate this aspect. We will also study the impact of inaccuracy in trace gathering. Here the work has been simply done with the help of HMM called as Hidden Markov Model but in future has include the high efficiency algorithms like MD5,AES,DES with good key values to provide the more security